\documentclass[conference, 9pt]{IEEEtran}
\IEEEoverridecommandlockouts
\usepackage[table,xcdraw]{xcolor}
\usepackage{comment}
\usepackage{amsmath,amssymb,amsfonts}
\usepackage{pifont}  

\usepackage{graphicx}
\usepackage[table,xcdraw]{xcolor}
\usepackage{textcomp}
\def\BibTeX{{\rm B\kern-.05em{\sc i\kern-.025em b}\kern-.08em
    T\kern-.1667em\lower.7ex\hbox{E}\kern-.125emX}}

\usepackage[super]{nth}
\usepackage{cite}
\usepackage[numbers,sort&compress]{natbib}
\usepackage{booktabs}
\usepackage{multirow}
\usepackage[ruled, lined, linesnumbered, commentsnumbered, noend]{algorithm2e}
\SetKwComment{Comment}{/* }{ */}
\usepackage{blindtext}
\usepackage{wrapfig}
\usepackage[caption=false]{subfig}
\usepackage[font=footnotesize]{caption}
\usepackage{tabularx}
\usepackage{soul}
\usepackage{tablefootnote}
\usepackage{enumitem}
\usepackage{algorithmic}
\usepackage{graphicx}
\usepackage{textcomp}
\usepackage{xcolor}
\def\BibTeX{{\rm B\kern-.05em{\sc i\kern-.025em b}\kern-.08em
    T\kern-.1667em\lower.7ex\hbox{E}\kern-.125emX}}

\newcommand{\cmark}{\textcolor{green!70!black}{\checkmark}}   
\newcommand{\xmark}{\textcolor{red}{\ensuremath{\times}}} 

\usepackage{tikz}
\newcommand{\circled}[1]{\tikz[baseline=(char.base)]{
            \node[shape=circle,draw=black,fill=black, text=white,inner sep=0.5pt] (char) {#1};}}
\begin{document}

\title{\emph{VeriOpt}: PPA-Aware High-Quality Verilog Generation via Multi-Role LLMs}
\author{Kimia Tasnia$^1$, Alexander Garcia$^1$, Tasnuva Farheen$^2$, Sazadur Rahman$^1$\\
\IEEEauthorblockA{
\textit{$^1$Department of ECE, University of Central Florida;$^2$Department of CSE, Lousiana State University} \\
\{$^1${kimia.tasnia, alexander.garcia, mohammad.rahman\}@ucf.edu; $^2$tfarheen@lsu.edu}}}

\maketitle

\begin{abstract}
The rapid adoption of large language models~(LLMs) in hardware design has primarily focused on generating functionally correct Verilog code, overlooking critical Power-Performance-Area~(PPA) metrics essential for industrial-grade designs. To bridge this gap, we propose VeriOpt, a novel framework that leverages role-based prompting and PPA-aware optimization to enable LLMs to produce high-quality, synthesizable Verilog. \emph{VeriOpt} structures LLM interactions into specialized roles (e.g., Planner, Programmer, Reviewer, Evaluator) to emulate human design workflows, while integrating PPA constraints directly into the prompting pipeline. By combining multi-modal feedback (e.g., synthesis reports, timing diagrams) with PPA aware prompting, \emph{VeriOpt} achieves PPA-efficient code generation without sacrificing functional correctness. Experimental results demonstrate up to $88\%$ reduction in power, $76\%$ reduction in area and $73\%$ improvement in timing closure compared to baseline LLM-generated RTL, validated using industry-standard EDA tools. At the same time achieves $86\%$ success rate in functionality evaluation. Our work advances the state-of-the-art AI-driven hardware design by addressing the critical gap between correctness and quality, paving the way for reliable LLM adoption in production workflows.
\end{abstract}

\begin{IEEEkeywords}
Design Automation, HDL, LLM, PPA Optimization
\end{IEEEkeywords}

\section{Introduction}


System on Chips (SoCs) contain multiple modules, including microprocessors, memory, and control logic. Hardware description languages (HDLs) such as Verilog, VHDL, SystemVerilog, or C++/SystemC are generally used to describe these modules. Electronic design automation processes, e.g., integration, synthesis, verification, placement, routing, and optimizations are then carried out on the HDL codes with target specifications, and technology libraries. The growing complexity and scale of SoCs, rapid advancement of semiconductor technology, and diminishing time-to-market windows, necessitate the implementation of automated tools and methodologies for HDL code generation to enhance the design flow, minimize human error, and expedite the entire design cycle from concept to silicon, thereby facilitating efficient design space exploration and power, performance, and area (PPA) optimized solutions. 

\begin{figure}[t]
    \centering
    \includegraphics[width=0.45\textwidth]{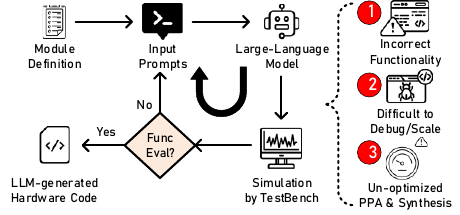}
    \caption{Existing LLM-based HDL generation flow and its limitations.}
    \label{fig:overview_limitation}
\end{figure}

Large language models (LLMs) are rapidly transforming HDL development by automating register-transfer-level (RTL) code generation~\cite{thakur2024verigen,thakur2023benchmarking,lu2024rtllm,chang2023chipgpt,liu2023rtlcoder, tasnia2025opl4gpt}, test bench writing~\cite{xu2024meic}, design optimization~\cite{chang2023chipgpt, pei2024betterv}, scripting of EDA tools~\cite{liu2023chipnemo}, resolving errors~\cite{tsai2023rtlfixer,liu2023chipnemo,qiu2024explaining,zhong2023llm4eda,yao2024rtlrewriter}, and analyzing reports~\cite{liu2023chipnemo}. Recent works leverage hierarchical decomposition~\cite{wei2025vflow, zhang2024mg, nakkab2024rome, sun2025paradigm}, multi-agent collaboration~\cite{wang2025rtlsquad, ho2024verilogcoder}, fine-tuning~\cite{chang2024data}, and chain-of-thought (CoT) techniques~\cite{yang2025haven, xu2024meic, tsai2023rtlfixer} to produce functionally correct RTL code, with commercial (e.g., GPT, Claude) and open-source (e.g., Llama) models adapted for hardware-specific tasks. As shown in Fig.~\ref{fig:overview_limitation}, existing LLM-based HDL generation approaches primarily focus on simulation based verification in-the-loop for syntactical and functional correctness. Despite their great promises, these methods produce partially functional codes~\cite{guo2025survey} which are difficult to debug and often overlooks scalability, interpretability, synthesizability, and most importantly, power, performance, and area (PPA) optimization, critical metrics for industrial-grade designs~\cite{fang2025survey, guo2025survey, chen2024large}. 

LLMs are language models trained with huge corpus of code repositories, such as, \emph{GitHub}~\cite{tasnia2025opl4gpt}, along with other modalities of datasets. The lack of diversity in \emph{GitHub} code distribution, specifically, the scarcity of HDL codes significantly limits their efficiency in RTL code generation~\cite{tasnia2025opl4gpt}.
Moreover, existing frameworks attempt to achieve functional and syntactical correctness through expensive fine-tuning, iterative refinement, or testbench validation~\cite{fang2025survey}, they lack systematic integration of hardware-specific optimization heuristics (e.g., resource sharing, timing closure). Without PPA awareness, LLM-generated RTL risks inefficiencies (e.g., redundant logic, suboptimal pipelining) that are costly or impossible to rectify during post-synthesis process~\cite{gandham2024circuitseer}. Therefore, it is critical to generate both functionally correct and PPA optimized code using LLMs. However, it is inherently challenging to generate PPA-optimized RTL code using LLMs due to fundamental gaps between language modeling and hardware design expertise pertaining from lack of PPA awareness in training data~\cite{akyash2025simeval}, design optimization aspects, trade-offs, and hardware physics.



To investigate the gap between LLM-generated RTL codes and realistic, optimized hardware codes, we explore the following research questions.

\noindent \underline{\textbf{RQ1.}} How can LLMs generate functionally correct code with transparency and readability without expensive pre-training or fine-tuning? 

\noindent \underline{\textbf{RQ2.}} Can LLM generate high-quality code considering different trade-offs of PPA metrics as per the design constraints?

By researching the above questions, in this paper we propose \emph{VeriOpt} by offering two key innovations: (i) role-based prompting and (ii) PPA-aware contextual learning. As shown in Fig.~\ref{fig:VeriOpt_framework}, unlike existing methods, \emph{VeriOpt} performs functional and PPA evaluation in the loop. The key contributions of this work is as follows.
\begin{itemize}[leftmargin=*]
    \item \emph{VeriOpt} addresses functional correctness via role-based framework that decomposes LLM tasks into four specialized roles - planner (breaks down the design into step-by-step sub-tasks), programmer (generates RTL code), reviewer (validates proper execution of the tasks), and evaluator (iteratively refines prompts based on error logs)—ensuring transparent, debug-friendly code generation. This systematic interplay of LLM role helps scalability, interpretability, and transparency for debugging.
    \item For PPA optimization, \emph{VeriOpt} injects hardware-specific expertise into prompts, guiding the LLM to apply power-saving techniques (clock-gating, power-gating, and resource sharing), area reduction (finite state machine encoding, resource sharing), and performance enhancements (pipelining, loop unrolling, and delay optimization). 
    \item By combining structured role interactions with domain-aware prompting, \emph{VeriOpt} achieves synthesizable, PPA-efficient RTL that aligns with industrial design constraints, surpassing LLM-generated `correct-but-suboptimal' outputs.
    \item \emph{VeriOpt} on RTLLM~\cite{lu2024rtllm} benchmark generates $86\%$ functionally correct code and enhances power, performance, and area of LLM-generated code by $88\%$, $73\%$, and $76\%$, respectively.
\end{itemize}

\section{Preliminaries and Motivation}
In this section, we first briefly review the existing works on LLM-based RTL generation and highlight their limitation. Later, we survey the established PPA optimization techniques in hardware codes before infusing out motivation behind \emph{VeriOpt}.

\subsection{Existing Works and Their Limitations}\label{subsec:existing_works}
\begin{table}[t]
\centering
\caption{Comparison of Recent works on RTL Generation by LLM. Here, PMT = Prompt Engineering, ICL = In-context learning}
\label{tab:background_comparison}
\resizebox{\columnwidth}{!}{%
\begin{tabular}{|c|cc|c|}
\hline
\multirow{2}{*}{\textbf{Works}} & \multicolumn{2}{c|}{\textbf{Objective}} & \multirow{2}{*}{\textbf{Approach}}\\ \cline{2-3} 
 &  \multicolumn{1}{c|}{Func.} & PPA & \\ \hline
Thakur et al.~\cite{thakur2023benchmarking, thakur2024verigen} & \multicolumn{1}{c|}{\checkmark} & $\times$ & Fine-tuning on golden RTL \\ \hline
Chang et al.~\cite{chang2023chipgpt} & \multicolumn{1}{c|}{\checkmark} & $\times$ & Prompt-based RTL \\ \hline
Lu et al~\cite{lu2024rtllm}& \multicolumn{1}{c|}{\checkmark} & $\times$ & Natural-language prompted \\ \hline
Liu et al.~\cite{liu2023rtlcoder} & \multicolumn{1}{c|}{\checkmark} & $\times$ & Fine-tuning with HDL\\ \hline
DeLorenzo et al.~\cite{delorenzo2024make} & \multicolumn{1}{c|}{\checkmark} & $\times$ & Monte Carlo search-guided \\ \hline
Pei et al.~\cite{pei2024betterv} & \multicolumn{1}{c|}{\checkmark} & $\times$ & Fine-tuning with HDL\\ \hline
Wang et al.~\cite{wang2025rtlsquad} & \multicolumn{1}{c|}{\checkmark} & $\times$ & Fine-tuning with HDL\\ \hline
Zhang et al.~\cite{zhang2024mg} & \multicolumn{1}{c|}{\checkmark} & $\times$ & Fine-tuning with ICL\\ \hline
Tang et al.~\cite{tang2024hivegen} & \multicolumn{1}{c|}{\checkmark} & $\times$ & Modular PMT\\ \hline
\textbf{This work (\emph{VeriOpt})} & \multicolumn{1}{c|}{\textbf{\checkmark}} & {\textbf{\checkmark}} & Role-based, PPA-aware PMT\\ \hline
\end{tabular}%
}
\end{table}
Recent advances in LLM-assisted hardware design have predominantly focused on achieving functional correctness, employing three main methodologies. 

\noindent\circled{1} \underline{Fine-Tuning with HDL Datasets:} Early adoption of LLM for RTL generation~\cite{thakur2023benchmarking, thakur2024verigen, liu2023rtlcoder, wang2025rtlsquad} utilize supervised fine-tuning (SFT) strategies on curated HDL datasets. While these methods significantly enhance syntax and functional correctness, they typically lack explicit optimization for critical hardware metrics such as power, performance, and area (PPA). Additionally, reliance on proprietary or high-quality datasets poses challenges for broad accessibility. 

\noindent\circled{2} \underline{Prompt Engineering (PMT):} Chang et al.~\cite{chang2023chipgpt} and Tang et al.~\cite{tang2024hivegen} apply structured prompting approaches, breaking down complex RTL generation tasks into manageable modules. Lu et al.~\cite{lu2024rtllm} advances further by leveraging natural-language-based prompt methods to simplify user interaction. Although highly effective for interactive and collaborative designs, these methods generally remain disconnected from hardware synthesis insights, thus lacking direct integration with PPA optimization processes. 

\noindent\circled{3} \underline{Search-Based and Hybrid Approaches:} DeLorenzo et al.~\cite{delorenzo2024make} introduced Monte Carlo Tree Search (MCTS), enhancing design space exploration and validation rigor. Zhang et al.~\cite{zhang2024mg} presented a hybrid approach combining supervised fine-tuning with in-context learning (ICL), improving generalization and error reduction in RTL generation tasks. Despite these innovative strategies, PPA metrics are usually addressed as post-design verification criteria rather than integrated optimization objectives. Moreover, post-LLM search imposes additional complexity to the RTL generation process. Table~\ref{tab:background_comparison} compares these methods, their objectives, and approach.

\subsection{PPA optimization techniques in Hardware Code Generation}\label{subsec:background_ppa}
\begin{figure}[t]
    \centering
    \includegraphics[width=0.5\textwidth]{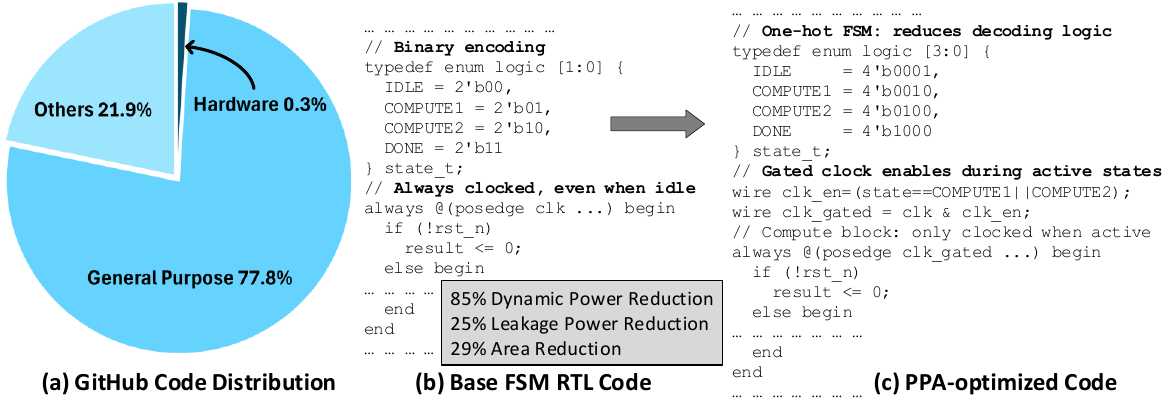}
    \caption{Challenges and limitations of current work. (a) Code distribution in GitHub, (b) an LLM-generated RTL code, and its (c) PPA-optimized version.}
    \label{fig:motivating_example}
\end{figure}
In real-world hardware design, teams often confront a delicate balancing act among three crucial hardware measures - power, performance, and area (PPA). Choosing to optimize for one usually imposes trade-offs on the others. For instance, clock gating or power gating can reduce dynamic and leakage power but may introduce additional area overhead and design complexity~\cite{weste2015cmos}. Similarly, pushing for higher clock frequency or deeper pipelines can boost performance but might lead to larger area and greater power consumption~\cite{hennessy2011computer}. Because every hardware application has different cost, speed, and energy constraints, it is essential for an LLM-based approach to generate Verilog codes that match the chosen priority, be it minimal power, minimal area, or maximum speed while adhering to functional requirements. In practice, area optimization often involves resource sharing (reducing dedicated hardware units and reusing them for multiple operations), dead code elimination, and careful fanout management to trim unnecessary buffering~\cite{weste2015cmos, hennessy2011computer}. Power optimization methods include clock gating and power gating (both disabling inactive circuitry), selecting low-power arithmetic operators, and minimizing switching activity through operand isolation~\cite{weste2015cmos}. Performance optimization, on the other hand, relies on architectural choices like pipelining (shortening critical paths)~\cite{hennessy2011computer}, retiming (balancing gate delays by moving registers)~\cite{leiserson1991retiming}, and loop unrolling (enabling parallel execution)~\cite{hennessy2011computer}. By systematically incorporating these techniques into RTL code, hardware developers can achieve designs that not only meet performance targets but also keep power and area in check—precisely the design adaptability that an LLM must offer in next-generation EDA workflows.

\subsection{Motivation}\label{subsec:background_motivation}
Fig~\ref{fig:motivating_example}(a) shows a distribution of GitHub codes for various programming languages. It can be noticed that hardware codes contributes on $0.3\%$ of the dataset~\cite{tasnia2025opl4gpt}. LLMs being trained with GitHub code repositories possess performance gap in Verilog code generation~\cite{fu2024generalize}. As highlighted in previous works, this data scarcity concern makes it challenging and expensive to bridge the gap by pretraining~\cite{guo2025survey}. Fine-tuning can be performed using relatively smaller HDL datasets, fewer optimization steps, and modest batch sizes~\cite{guo2025survey}. However, as discussed in Sec~\ref{subsec:existing_works}, the lack of high-quality, diverse, and open-source hardware dataset makes fine-tuned models to under-perform~\cite{thakur2023benchmarking, thakur2024verigen, liu2023rtlcoder, wang2025rtlsquad}. Therefore, as specified in \textbf{RQ1}, in this work, our motivation is to explore multi-role LLM-based prompting, as an alternative, to generate functionally correct Verilog codes.

Although achieving functional correctness in RTL is crucial, omitting PPA metrics at this early design phase can have grave repercussions later~\cite{guo2025survey, fang2025survey}. Let us consider a System-on-Chip (SoC) team that only identifies massive power leakage after tape-out, discovering that the idle blocks were not properly clock-gated. Rectifying this design flaw means a multi-million-dollar re-spin of the whole design, implementation, and manufacturing cycle. Worse still, if performance issues manifest like failing to close timing at the desired clock frequency, the entire product launch can be delayed or forced to operate at suboptimal speeds. Fig.~\ref{fig:motivating_example}(b-c) illustrates how seemingly minor oversights in low-level RTL eventually amplify into large-scale cost, time, and performance penalties downstream. By rewriting the baseline Verilog code of Fig.~\ref{fig:motivating_example}(b) for one-hot encoding and clock-gating, one can reduce the dynamic power, leakage power, and area by $85\%$, $25\%$, and $29\%$, respectively. Therefore, \emph{VeriOpt} addresses \textbf{RQ2} by injecting PPA-oriented context into LLMs, enabling them to produce Verilog codes optimized for power, performance, and area.

Motivated by these high stakes, \emph{VeriOpt} fuses role-based prompting with PPA-aware in-context learning to pursue both (i) functional correctness and (ii) design quality. Essentially, we aim for LLMs that generate not only syntactically and semantically valid code, but also code that meets PPA criteria established by hardware engineers. We detail these methods in Sec.~\ref{sec:methodology}.

\section{Methodology}\label{sec:methodology}
In this section we present the \emph{VeriOpt} framework by explaining its role-based prompting and PPA-aware in-context learning mechanism.
\begin{figure}[t]
    \centering
    \includegraphics[width=0.5\textwidth]{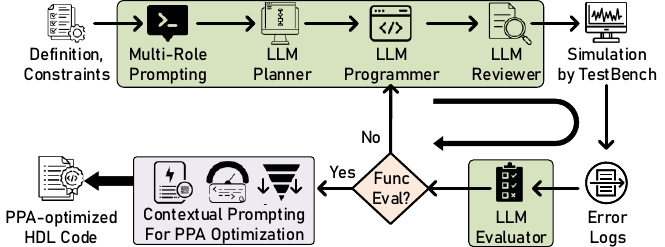}
    \caption{Overview of \emph{VeriOpt} LLM-based HDL coding flow with multi-role (green) and contextual PPA-aware prompting (purple).}
    \label{fig:VeriOpt_framework}
\end{figure}

\begin{figure*}[t]
    \centering
    \includegraphics[width=1.0\textwidth]{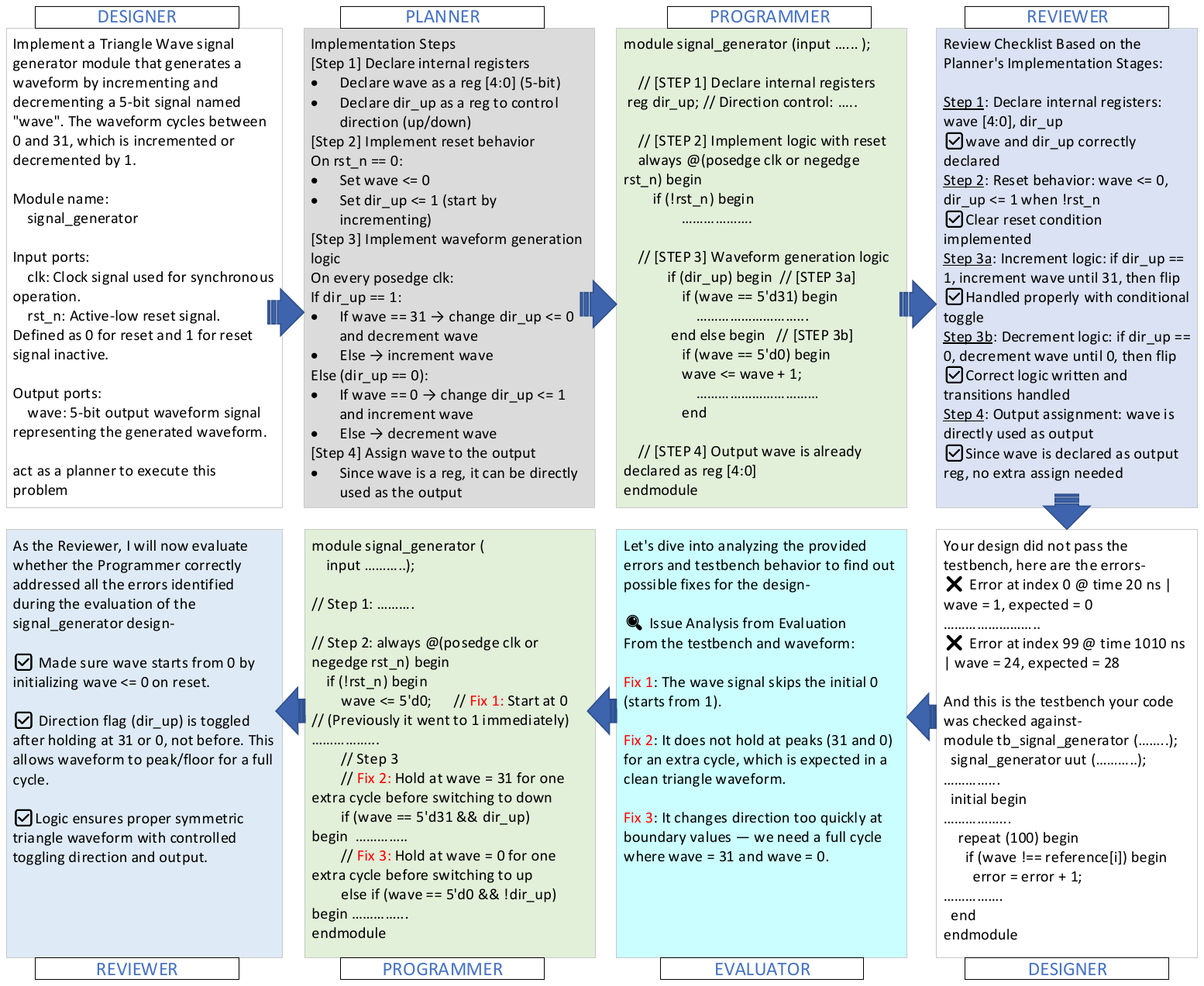}
    \caption{Multi role based LLM's (Planner, Programmer, Reviewer, Evaluator) flow diagram to develop a functionally correct Verilog code.}
    \label{fig:multi_role}
\end{figure*}


\subsection{RQ1: How LLM Can Generate Functionally Correct Code}\label{subsec:multi_role_LLM}
In this subsection we first emphasize why existing LLM-based HDL coding methods fails to generate functionally correct code due to surface-level modification and lack of reasoning. Later, we explain the multi-role prompting mechanism proposed by \emph{VeriOpt} that mimics human coding flow.

\subsubsection{Challenges in Functional Code Generation}
Most existing works treat large language models (LLMs) as black-boxes modifying only surface-level components, such as, the fine-tuning/training dataset~\cite{thakur2023benchmarking, thakur2024verigen, liu2023rtlcoder, wang2025rtlsquad}, prompt template~\cite{chang2023chipgpt, tang2024hivegen, lu2024rtllm}, or post-processing search strategy~\cite{delorenzo2024make, zhang2024mg} without investigating how the LLM internalizes and reasons through hardware design tasks. This oversight leaves a critical gap regarding intermediate cognitive steps required for reliable HDL generation - \textbf{how LLMs decompose} natural language into hardware primitives (e.g., registers, FSMs), \textbf{why they select} specific operators or coding styles (e.g., blocking vs. non-blocking assignments), \textbf{where they struggle} to align high-level intent with synthesizable RTL constructs. Without exposing these intermediate steps, it becomes extremely difficult (iterative process) to trace the source of functionality mismatches or to debug flawed RTL behavior. Moreover, these intermediate steps serve as a chain-of-thought mechanism, which is empirically known to improve reasoning quality in code generation tasks~\cite{wei2022chain}. In the next subsection, we discuss, how \emph{VeriOpt} addresses this gap through a multi-role prompting framework that decomposes HDL generation into transparent, intermediate stages - planning, programming, reviewing, and evaluating, thus offering interpretability, correctness-by-construction, and guided debugging throughout the RTL development flow mirroring human workflow. The documentation of intermediate decisions enables stepwise validation, and embeds hardware-aware checks before code completion.

\subsubsection{Multi-Role Prompting}
As discussed in previous subsection, we propose a multi-role based prompting techniques to address the reasoning and transparency concern of LLM-based HDL generation. In this approach, LLM plays different roles as a - \textbf{\underline{Planner}}, \textbf{\underline{Programmer}}, \textbf{\underline{Reviewer}} and \textbf{\underline{Evaluator}} to generate a functionally correct RTL code as presented in Fig~\ref{fig:VeriOpt_framework}. In this figure, first design specifications are provided to the \textit{LLM Planner} to plan the execution strategy. The \textit{LLM Programmer} follows the planning steps and develops the Verilog code. The developed code is reviewed by the \textit{LLM Reviewer} to ensure all the planning strategies have been executed in the code. Followed by, the \textit{LLM Evaluator} assesses and analyzes the code for any potential errors and suggests possible fixes. Fixing and evaluating goes on a feedback process until a functionally correct code is generated. The details of this framework with corresponding examples are described in the following sections.

\vspace{0.1in}
\noindent \textbf{LLM Planner:}
A good planning is a prerequisite for any successful task. Similarly for RTL, designers plan how they are going to develop the code according to the specification. Being inspired from this, the first role of LLM is a planner to layout the execution steps based on the given problem statement as follows.
\begin{itemize}[leftmargin=*]
    \item LLM is given the problem as a plain text with module definitions - module name, and input/output ports based on the specification. 
    \item LLM as a planner creates a number of implementation steps that describe every granular stage of the code it plans to execute.
    \item The steps include low-level hardware primitives, such as, any internal registers and parameters necessary to define, states for finite state machine (FSM), number of always blocks, blocking or non-blocking assignments, and so on. 
\end{itemize}
These implementation steps help to get a clear understanding of LLM`s thought process as a planner to the designer to address the problem statement. In Fig.~\ref{fig:multi_role} a functionally correct `Signal Generator' RTL code development flow is presented involving different roles of LLMs. The \textit{LLM planner} planned $4$ implementation steps to execute based on the given problem statement from the designer. Here, we provided this problem with input and output ports as a RTL designer.

\vspace{0.1in}
\noindent \textbf{LLM Programmer:}
As shown in Fig.~\ref{fig:VeriOpt_framework}, followed by the planning stage, LLM plays the role as a programmer with following sub-tasks.
\begin{itemize}[leftmargin=*]
    \item The programmer develops the Verilog code as per the description of the implementation steps. It executes exactly each step planned by the planner while developing the code. 
    \item It also adds comments alongside the code lines, explicitly showing the execution of the implementation steps which later helps as a reference of reasoning for debugging. 
\end{itemize}
In this way, the designer can follow whether the LLM executed every step it previously planned out and did not change the implementation flow out of the designer's knowledge. Fig.~\ref{fig:multi_role} shows the code developed by LLM Programmer for `Signal Generator' followed by the \textit{LLM planner}'s implementation steps. Each step is explicitly mentioned with comment inside the code.

\vspace{0.1in}
\noindent \textbf{LLM Reviewer:}
To verify whether LLM is following the exact steps, at this stage we guide it to act as a reviewer with following tasks.
\begin{itemize}[leftmargin=*]
    \item Reviewer will ensure the Programmer executed all the implementation steps in it's code. 
    \item It will check the generated RTL thoroughly and explain each implementation step addressed inside the codes. 
    \item With it's careful reviewing, it will find out if any step is missed by the programmer and indicate so that the programmer can fix it. 
\end{itemize}
With this step, \textit{LLM Reviewer} not only confirms the proper execution of the implementation steps but also provide the transparency to the designer in the code generated by LLM. Fig.~\ref{fig:multi_role} shows how the \textit{LLM Reviewer} verified each and every implementation steps in the RTL code and labeled with check marks for successful implementation. 

\vspace{0.1in}
\noindent \textbf{LLM Evaluator:}
After successfully confirming that all the implementation steps are executed by the \textit{LLM Programmer}, the next role is to act as an evaluator as follow to verify the functional correctness.
\begin{itemize}[leftmargin=*]
    \item Evaluator verifies the syntactical and functional correctness of the code developed by LLM Programmer.
    \item The flow also runs the \textit{LLM Programmer}-generated Verilog code with the corresponding testbench in the verification tool. If any error message shows up, the error logs along with the corresponding testbench is provided to the LLM evaluator as a feedback. 
    \item \textit{LLM Evaluator} critically assesses the errors and testbench to indicate the possible fixes to the programmer. 
    \item Programmer rewrites the code by adding explicit comments to indicate where it has performed the fixes. 
    \item The updated code goes through the review process by the \textit{LLM Reviewer} to ensure all the error fixes analyzed by \textit{LLM Evaluator} have been addressed. 
    \item The fixed code then goes through the same feedback loop until the successful execution of a functionally correct code. 
\end{itemize}
Adding comments to the fixed code by \textit{LLM Programmer} and \textit{LLM Reviewer}'s confirmation of the fixes inside the code clearly shows a transparent workflow. Therefore, the designer no longer needs to blindly rely on LLM's job of evaluating and rewriting the code; rather, he/she can monitor each step easily to find any ambiguity.

Fig.~\ref{fig:multi_role} presents these steps followed by the \textit{LLM Reviewer}'s role. After getting several errors while running against the `Signal Generator' testbench, we provided these error logs along with the testbench to the \textit{LLM Evaluator}. The evaluator analyzed these errors and suggested $3$ possible fixes. In the following step, \textit{LLM Programmer} implemented these fixes and explicitly specified the fixes with comments alongside the code lines. Followed by, the \textit{LLM reviewer} confirmed that all these fixes have been addressed by the \textit{LLM Programmer}. This fixed code again went through the same evaluation process until passed successfully in the second iteration. Sec.~\ref{subsec:rq1_results} and Table~\ref{tab:comparison_func} show the application of multi-role prompting by \emph{VeriOpt} on the RTLLM~\cite{lu2024rtllm} benchmark to generate functionally correct codes.

\subsection{RQ2: PPA-Aware In-Context Learning}
\begin{figure}[t]
    \centering
    \includegraphics[width=0.45\textwidth]{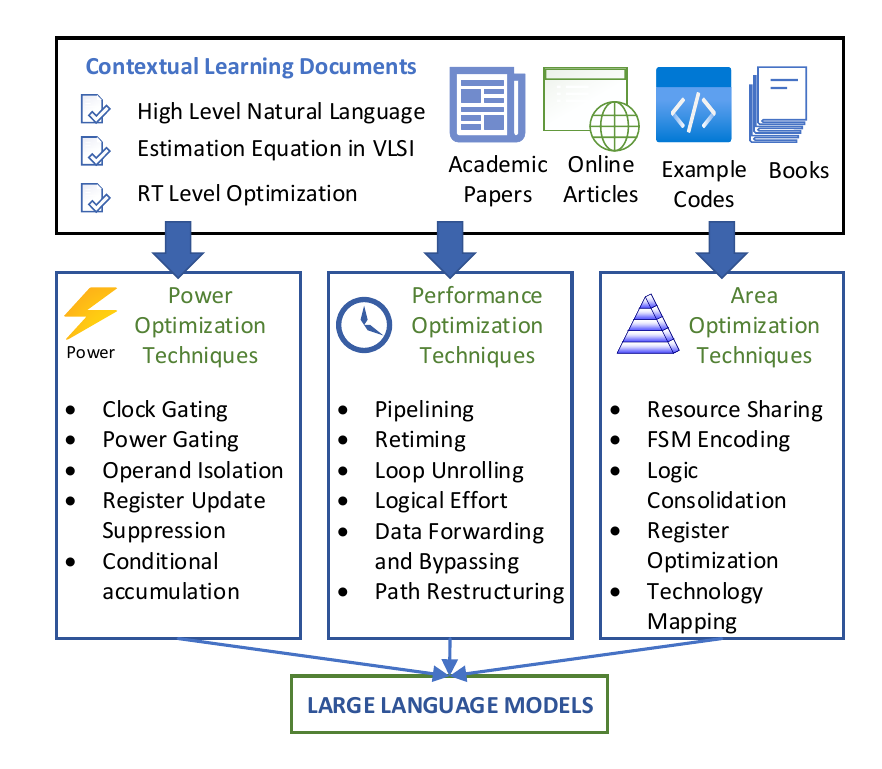}
    \caption{In-context learning for PPA optimization by \emph{VeriOpt}. We leverage domain-specific knowledge~\cite{weste2015cmos}—including power, performance, and delay estimation equations, PPA optimization strategies, and example code snippets~\cite{openmsp}—to infuse PPA-awareness into the LLM.}
    \label{fig:ICL}
\end{figure}  

\emph{VeriOpt} enables designer trust in the AI-generated code. While naive code-generation solutions might produce `correct' answers that contain hidden or extraneous logic (i.e., redundant/suboptimal code), our system fosters quality Verilog generation by optimizing for PPA. The following subsections discuss challenges with generating PPA-aware code generation by LLM and how our \emph{VeriOpt} address this.

\subsubsection{Challenges in Fusing PPA-awareness to LLM}\label{PPA Challenges}
As discussed in Sec.~\ref{subsec:existing_works} and ~\ref{subsec:background_motivation}, enabling LLMs for PPA-aware code generation remains non-trivial. Fundamentally, LLMs are trained as sequence predictors and lack intrinsic reasoning or domain-specific optimization capabilities unless explicitly guided. To fully appreciate the problem space, we first identify the challenges in building PPA-aware Verilog code generation capability within LLM. 

\noindent\circled{C1} LLMs lack intrinsic capabilities to quantify PPA metrics from code alone~\cite{guo2025survey, fang2025survey}. Unlike human designers who interpret synthesis reports or tool feedback, LLMs operate purely on textual patterns.

\noindent\circled{C2} LLMs must pinpoint opportunities for improvement by analyzing the structure of the generated code~\cite{guo2025survey, fang2025survey}, e.g., deciding whether clock gating or resource sharing is beneficial without compromising the functionality. As discussed in Sec.~\ref{subsec:background_ppa}, context-dependent analysis is critical as each design may require a combination of techniques.

\noindent\circled{C3} Optimizing PPA at RTL is inherently difficult~\cite{gandham2024circuitseer} due to the lack of physical implementation details, such as, routing delays, cell sizes, whereas, LLMs can only generate behavioral RTL codes.

\noindent\circled{C4} LLMs must apply RTL optimization paradigms discussed in Sec.~\ref{subsec:background_ppa}, such as, pipelining, loop unrolling, or low-power arithmetic in a way that balances multiple objectives~\cite{li2024s}. No single technique is universally optimal across all hardware modules, which demands nuanced decision-making about transformations that serve best. 

\noindent\circled{C5} Designers express PPA intent in high-level natural language (e.g., minimize power while meeting 500MHz timing). Translating this to RTL requires mapping abstract constraints to microarchitectural transformations (e.g., power gating, resource sharing), preserving correctness while applying optimizations, a task demanding hardware-specific reasoning beyond token prediction~\cite{chen2024large}. 

The following Sec.~\ref{subsec:veriopt_ppa} explains how \emph{VeriOpt} addresses these challenges by contextual learning.

\subsubsection{Contextual Learning Strategies by \emph{VeriOpt}}\label{subsec:veriopt_ppa}

In-context learning is a way to guide language models to learn tasks given only a few examples allowing users to quickly build models for a new use case without going through expensive and time-consuming fine-tuning and storing new parameters for each task~\cite{li2023large}. It typically requires very few training examples to get a prototype working, and the natural language interface is intuitive even for non-experts. Several challenges are imposed while building PPA-aware domain-specific capabilities in developing Verilog code as described in Sec.~\ref{PPA Challenges}. \emph{VeriOpt} addressed these challenges in its contextual learning strategies as follows.

\noindent\circled{S1} \emph{VeriOpt} has chosen the in-context learning (ICL) contents such a way that those are high-level natural languages easy for LLM to understand. For example, estimating code timing delays by static timing analysis methods (using technology libraries and timing graphs) would have been quite difficult for LLM to evaluate the RTL code performance. However, \emph{VeriOpt} provides timing context by infusing the fundamentals of parasitic delay estimation techniques regularly followed in VLSI concepts~\cite{weste2015cmos}, making it easier for LLM to comprehend. Nonetheless, all the PPA enhancement techniques utilized is ICL, are of architectural or coding paradigms based at RT-level. For example, pipelining is an RT-level exclusive performance enhancement approach which can not be applied in the gate-level.  

\noindent\circled{S2} PPA optimization has trade-offs in balancing power, performance and area metrics. Based on the cost, speed and energy constraints of different hardware designs, it is crucial for LLMs to be aware of different optimization techniques, their scopes, and constraints for various RTL design structure. To address this challenge, we chose several PPA optimization techniques, as shown in Fig.~\ref{fig:ICL}, to meet specific design goal. Our techniques include but not limited to -

\noindent\textbf{Power optimized:} Clock gating, power Gating, operand isolation, register update suppression, conditional accumulation.

\noindent\textbf{Performance optimized:} Pipelining, register retiming, loop unrolling, logical effort, path restructuring.

\noindent\textbf{Area optimized:} Resource sharing, FSM state encoding, logic consolidation, register optimization, technology mapping.

For each technique, we provide LLM relevant content from literature, such as, VLSI books~\cite{weste2015cmos}, corresponding examples in Verilog codes~\cite{openmsp, loopunrolling}, online available articles from different vendors including real-life application~\cite{vivadosynth, powergating} and academic papers of different authors~\cite{fsmlowpower, clockgating, loopoptimize, stateencoding}.

\noindent\circled{S3} Since LLMs lack the capability to quantify PPA metrics from the Verilog code alone, in \emph{VeriOpt} we provide the synthesis report of the base Verilog code that we intend to optimize. The synthesis report consists of the following design metrics - (\underline{\textbf{power}}) the dynamic and leakage power, cell internal and net switching power; (\underline{\textbf{area}}) cell area and design area, combinational and sequential area; (\underline{\textbf{timing}}) critical path length, critical path slack, total negative slack and levels of logic. By analyzing these synthesis metrics, LLM interprets the PPA efficiency of the base design and the possible scopes for improvement.  
\begin{figure*}[t]
    \centering
    \includegraphics[width=1.0\textwidth]{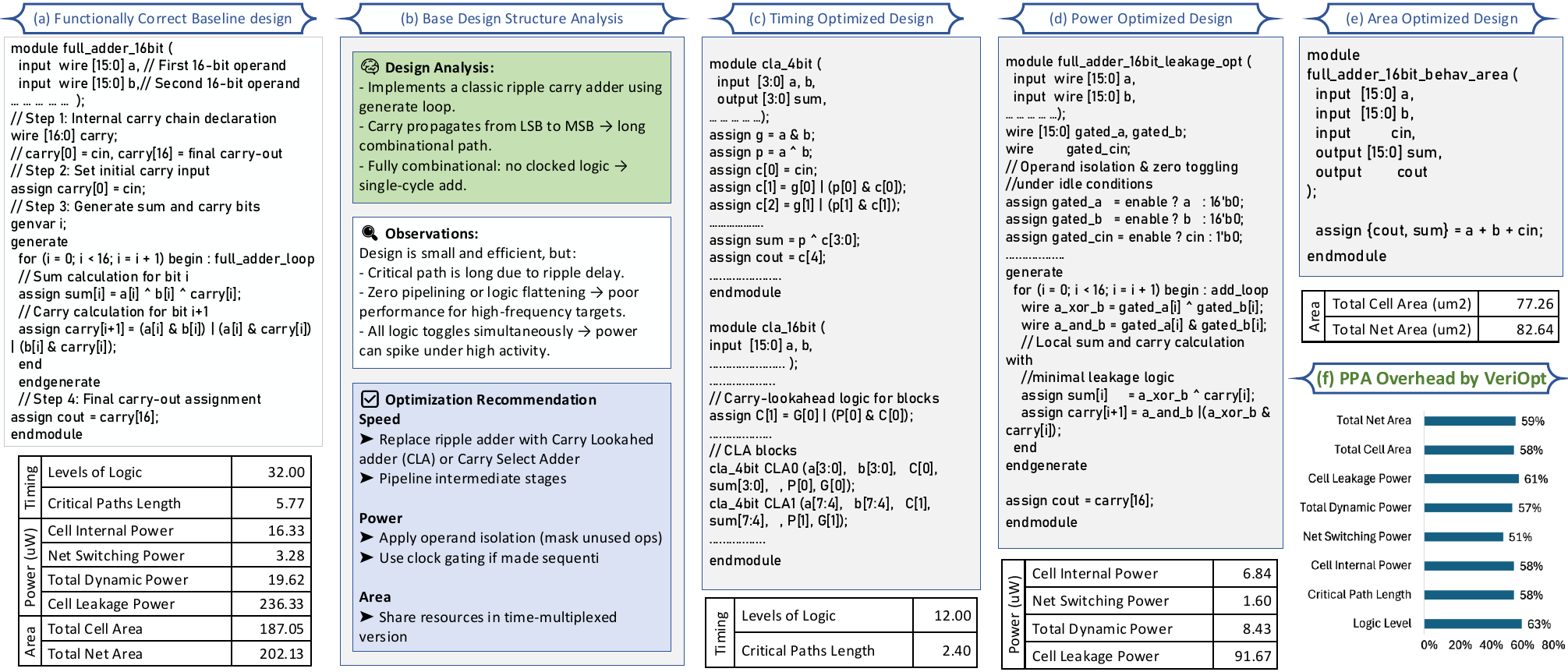}
    \caption{\emph{VeriOpt} generated PPA optimized designs. (a) A functionally correct but PPA un-optimized baseline design developed by multi-role LLM from Sec.~\ref{subsec:multi_role_LLM}, (b) \emph{VeriOpt} analyzes the design structure, synthesis report and suggests PPA optimization techniques, (c) timing, (d) power, (e) area optimized design along with their corresponding synthesis reports, (f) PPA improvement by \emph{VeriOpt} across all the metrics.}
    \label{fig:PPA optimized}
\end{figure*}

\noindent\circled{S4} With all the comprehendible knowledge of various PPA techniques and base design synthesis metric reports, \emph{VeriOpt} can efficiently analyze the design to choose the best optimization strategies. First it analyzes the RTL structure, whether it is a fully combinational circuit or sequential, type of sequential block (e.g. iterative shift and add operation), reset structure, core logic involved in the code, number of operands, registers, control signals, any pipelining stages etc. Based on the understanding of the RTL structure, it suggests the best applicable techniques to optimize power consumption, timing delay and area utilization. \emph{VeriOpt} can also adopt a combination of different PPA optimization techniques to get the best possible result. 

Fig.~\ref{fig:ICL} presents the summary of \emph{VeriOpt}'s in-context learning strategies, providing selected RT-level optimization techniques in terms of power, performance, and area with documents from VLSI books~\cite{weste2015cmos}, online articles~\cite{vivadosynth, powergating}, academic paper~\cite{loopoptimize, fsmlowpower, stateencoding, clockgating} with corresponding example codes in Verilog~\cite{openmsp, loopunrolling}. All these contents are documented in high-level natural language, with estimation equations in scope for RT-level. 

\subsubsection{A Case Study with \emph{VeriOpt} PPA Optimization}\label{case quality}
Fig.~\ref{fig:PPA optimized} shows a practical application of the PPA optimization technique on a full $16$-bit adder code from RTLLM benchmark~\cite{lu2024rtllm} utilizing \emph{VeriOpt}. The baseline Verilog code and optimized code in terms of power, timing and area is presented in the figure for providing a clear picture of the implementation outcomes.
\begin{itemize}[leftmargin=*]
    \item First, we provide the in-context learned LLM with the functionally correct baseline code of the full $16$-bit adder which was designed by the multi-role LLMs framework described in Sec.~\ref{subsec:multi_role_LLM}.
    \item Along with the Verilog code, the synthesis report consisting of power, timing and area metrics are provided  as shown in Fig.~\ref{fig:PPA optimized}(a). It helps \emph{VeriOpt} to better understand the PPA status of the functionally correct but not optimal baseline code. 
    \item Next, \emph{VeriOpt} carefully analyzes the RTL code structure and the synthesis report. Based on the analysis, it illustrates its observation of the design inefficiencies and possible scopes of improvement in terms of power, performance, and area as shown in Fig.~\ref{fig:PPA optimized}(b). 
    \item After identifying the design's structure and inefficiencies, \emph{VeriOpt} recommends different enhancement techniques to optimize for power, timing delay and area. It can be noticed from Fig.~\ref{fig:PPA optimized}(b), for this full $16$-bit adder design, \emph{VeriOpt} recommends - operand isolation and clock gating technique for \underline{\textbf{power}} optimization; replacing ripple adder with carry look ahead adder and pipelining for \underline{\textbf{timing}} optimization; and resource sharing for \underline{\textbf{area}} optimization. 
    \item By applying these techniques suggested by \emph{VeriOpt}, we can generate three different - power-optimized, timing-optimized, and area-optimized designs as shown in Fig.~\ref{fig:PPA optimized}(c), (d), and (e), respectively along with their synthesis reports in the bottom. The PPA overhead ($\%$) achieved by \emph{VeriOpt} is also presented in a graph in Fig.~\ref{fig:PPA optimized}(f) showing the significant reduction in all the corresponding metrics - $58.7\%$ cell area, $59.11\%$ design area, $57.03\%$ dynamic power, $61.21\%$ leakage power and $58.4\%$ critical path length.
\end{itemize} 
\noindent In Sec.~\ref{subsec:rq2_results} and Table~\ref{tab:ppa_comparison} of the following section, we apply this in-context learning-based PPA-aware prompting approach of \emph{VeriOpt} on the widely accepted RTLLM~\cite{lu2024rtllm} benchmarks to understand its effectiveness in generating optimal Verilog codes.

\section{Experimental Results}\label{sec:results}
In this section, we present the experimental setup, evaluation metrics, and perform functional and PPA evaluation of \emph{VeriOpt}.
\vspace{-0.1in}
\subsection{Experimental Setup}
\subsubsection{Baselines and Benchmarks} In this study, we use RTLLM~\cite{lu2024rtllm} benchmark as our evaluation dataset which contains $29$ designs of various categories and complexity levels. \emph{VeriOpt}'s performance in functional Verilog code generation is compared with other methods, such as, Thakur et. al.~\cite{thakur2023benchmarking}, RTLCoder~\cite{liu2023rtlcoder} and RTLLM's Self-Planning~\cite{lu2024rtllm}. We also present some other baseline LLM results of GPT-3.5, GPT-4, and StarCoder. 

\subsubsection{Evaluation Metrics} We focus on two main evaluation metrics based on our two research questions we emphasized previously. 

\noindent\textbf{RQ1:} For evaluating the functional success rate of \emph{VeriOpt} generated RTL code, first we check the syntactical correctness of the code. The syntax-wise correct codes are evaluated against RTLLM's~\cite{lu2024rtllm} golden testbench for functionality test. The designs successfully meeting all the test cases of RTLLM's testbench are considered as functionally `pass' otherwise a `fail'. All the other benchmarks and baselines LLMs are evaluated in the same criteria for a fair evaluation.

\noindent\textbf{RQ2:} With a view to developing high quality RTL code, we evaluate PPA optimized designs by \emph{VeriOpt} with the functionally correct base design in the previous step before optimization. We individually evaluate every design for power, performance and area enhancement compared to the base design. From the synthesis report, we mainly focus on dynamic power, leakage power, critical path length, critical path slack, cell area, and design area as PPA metrics for evaluation.

\begin{table*}[t]
\centering
\caption{Comparison on functionally correct code generation by existing works and \emph\{VeriOpt\} on RTLLM benchmarks~\cite{lu2024rtllm} with success rate of each method mentioned in the bottom row. The functionally passed cases are marked $\cmark$ and $\xmark$ otherwise. The syntactically failed cases are marked ``-".}
\label{tab:comparison_func}
\setlength{\tabcolsep}{0.5pt}
\resizebox{\linewidth}{!}{%
\begin{tabular}{l|c|c|c|c|c|c|c|c|c}
\hline
\textbf{Design} & \textbf{GPT-3.5} & \textbf{GPT-4} & \textbf{Thakur et al.} & \textbf{StarCoder} & \textbf{RTLCoder-Mistral-4bit} & \textbf{RTLCoder-Mistral} & \textbf{RTLLM GPT-3.5} & \textbf{RTLLM GPT-4} & \textbf{VeriOpt} \\ \hline
accu & \cmark & \cmark & - & - & $\xmark$ & $\xmark$ & \cmark & \cmark & \cmark \\
adder 8bit & \cmark & \cmark & - & - & \cmark & \cmark & \cmark & \cmark & \cmark \\
adder 16bit & $\xmark$ & \cmark & \cmark & - & - & $\xmark$ & \cmark & \cmark & \cmark \\
adder 32bit & $\xmark$ & $\xmark$ & - & - & $\xmark$ & $\xmark$ & \cmark & $\xmark$ & \cmark \\
adder 64bit & $\xmark$ & $\xmark$ & - & - & $\xmark$ & $\xmark$ & $\xmark$ & $\xmark$ & \cmark \\
multi 8bit & $\xmark$ & $\xmark$ & - & - & \cmark & \cmark & $\xmark$ & $\xmark$ & \cmark \\
multi 16bit & - & \cmark & \cmark & - & $\xmark$ & \cmark & \cmark & \cmark & \cmark \\
multi pipe 4bit & - & \cmark & - & \cmark & $\xmark$ & $\xmark$ & $\xmark$ & $\xmark$ & \cmark \\
multi pipe 8bit & - & $\xmark$ & - & \cmark & - & $\xmark$ & $\xmark$ & \cmark & \cmark \\
div 8bit & - & - & - & - & $\xmark$ & $\xmark$ & $\xmark$ & - & $\xmark$ \\
div 16bit & - & $\xmark$ & - & - & - & - & \cmark & $\xmark$ & \cmark \\
JC counter & \cmark & \cmark & $\xmark$ & $\xmark$ & \cmark & \cmark & \cmark & \cmark & \cmark \\
right shifter & \cmark & \cmark & \cmark & - & \cmark & \cmark & - & \cmark & \cmark \\
mux & - & \cmark & \cmark & - & \cmark & \cmark & $\xmark$ & \cmark & \cmark \\
counter 12 & \cmark & \cmark & $\xmark$ & $\xmark$ & \cmark & \cmark & \cmark & \cmark & \cmark \\
freq div & $\xmark$ & $\xmark$ & $\xmark$ & - & \cmark & \cmark & $\xmark$ & \cmark & \cmark \\
signal generator & \cmark & \cmark & - & $\xmark$ & $\xmark$ & $\xmark$ & \cmark & \cmark & \cmark \\
serial2parallel & \cmark & \cmark & - & $\xmark$ & $\xmark$ & $\xmark$ & \cmark & \cmark & \cmark \\
parallel2serial & $\xmark$ & $\xmark$ & - & - & $\xmark$ & \cmark & $\xmark$ & $\xmark$ & $\xmark$ \\
pulse detect & $\xmark$ & $\xmark$ & $\xmark$ & - & $\xmark$ & $\xmark$ & $\xmark$ & $\xmark$ & $\xmark$ \\
edge detect & \cmark & \cmark & $\xmark$ & - & \cmark & \cmark & \cmark & \cmark & \cmark \\
FSM & $\xmark$ & $\xmark$ & $\xmark$ & - & $\xmark$ & $\xmark$ & $\xmark$ & $\xmark$ & \cmark \\
width 8to16 & \cmark & \cmark & - & \cmark & \cmark & \cmark & \cmark & \cmark & \cmark \\
traffic light & $\xmark$ & \cmark & - & - & \cmark & \cmark & $\xmark$ & \cmark & $\xmark$ \\
calendar & - & $\xmark$ & - & - & $\xmark$ & $\xmark$ & \cmark & \cmark & \cmark \\
RAM & - & - & \cmark & \cmark & \cmark & \cmark & - & \cmark & \cmark \\
asyn fifo & - & - & - & - & - & $\xmark$ & $\xmark$ & - & \cmark \\
ALU & - & $\xmark$ & - & - & $\xmark$ & $\xmark$ & - & \cmark & \cmark \\
PE & \cmark & \cmark & $\xmark$ & \cmark & \cmark & \cmark & \cmark & \cmark & \cmark \\ \midrule
Success Rate & 10/29 & 15/29 & 5/29 & 5/29 & 12/29 & 14/29 & 14/29 & 19/29 &  \textbf{25/29}\\
\bottomrule
\end{tabular}%
}
\end{table*}

\subsubsection{LLM and EDA Tools} \emph{VeriOpt} leverages OpenAI's state-of-thre-art~(SOTA) GPT-4o model for multi-role prompting and in-context learning based PPA optimization. For syntax and functionality verification, we use Icarus Verilog EDA tool. To synthesize designs for PPA metrics evaluation, we employ Synopsys Design Compiler.

\vspace{-0.1in}

\subsection{RQ1: Functionality Evaluation and Analysis}\label{subsec:rq1_results}
Table~\ref{tab:comparison_func} presents the functionality evaluation of \emph{VeriOpt}'s multi-role prompting technique. Our experimental result shows that \emph{VeriOpt} successfully passed $25$ out of $29$ test scenarios of the RTLLM benchmark. Our framework clearly outperforms baseline LLMs (GPT-4 passed 15 out of 29), and other LLM-based Verilog code generation methods, such as, RTLCoder~\cite{liu2023rtlcoder} (12/29), Thakur at al.~\cite{thakur2023benchmarking} (5/29), and RTLLM's Self-Planning~\cite{lu2024rtllm} (19/29). Also \emph{VeriOpt} generates $100\%$ syntactically correct codes. The result indicates that \emph{VeriOpt} - 1) can successfully generate functionally correct code outperforming the existing methods with the highest success rate (RTLLM's Self-Planning~\cite{lu2024rtllm}), 2) possesses the robustness of decomposing natural language description into comprehendible implementation steps without even involving expensive pre-training or fine-tuning~\cite{thakur2023benchmarking, thakur2024verigen, liu2023rtlcoder, wang2025rtlsquad}. The debugging in feedback loop by fixing one/two implementation step only with explicit comments in the code makes it achievable and successful. These results reflect our findings for \textbf{RQ1} and efficiency of \emph{VeriOpt} in functional Verilog code generation.



\subsection{RQ2: Quality Evaluation and Analysis}\label{subsec:rq2_results}
\begin{table*}[t]
\centering
\caption{Comparison of PPA enhancement by \emph{VeriOpt} on RTLLM Benchmark. A positive value (green) for \emph{VeriOpt} represents improvement here, where a negative value (red) represents the opposite. In cases where the RTLLM benchmark is a combinational design, timing slack is marked a `N/A - Not Applicable'.}
\label{tab:ppa_comparison}
\resizebox{\textwidth}{!}{%
\begin{tabular}{@{}l|rr|rr|rr|rr|rr|rr@{}}
\toprule
\multicolumn{1}{c}{\multirow{3}{*}{Design}} & \multicolumn{6}{c}{Baseline} & \multicolumn{6}{c}{\emph{VeriOpt} Optimized (\% Improved)} \\
\cmidrule(lr){2-7}\cmidrule(lr){8-13}
\multicolumn{1}{c}{} & \multicolumn{2}{c}{Area (µm2)} & \multicolumn{2}{c}{Power (µW)} & \multicolumn{2}{c}{Timing (ns)} & \multicolumn{2}{c}{Area (\%)} & \multicolumn{2}{c}{Power (\%)} & \multicolumn{2}{c}{Timing (\%)} \\
\cmidrule(lr){2-3}\cmidrule(lr){4-5}\cmidrule(lr){6-7}\cmidrule(lr){8-9}\cmidrule(lr){10-11}\cmidrule(lr){12-13}
\multicolumn{1}{c}{} & Cell & Design & Dynamic & Leakage & CP length & CP Slack & Cell & Design & Dynamic & Leakage & CP length & CP Slack \\
\midrule
accu & 510.07 & 580.28 & 118.82 & 0.78 & 1.79 & -1.5 & \color[HTML]{32CB00}9.12 & \color[HTML]{32CB00}7.79 & \color[HTML]{32CB00}6.69 & \color[HTML]{32CB00}28.1 & \color[HTML]{32CB00}12.84 & \color[HTML]{32CB00}16.66 \\
adder 8bit & 89.46 & 96.12 & 8.24 & 0.07 & 1.9 & N/A & \color[HTML]{32CB00}56.82 & \color[HTML]{32CB00}56.97 & \color[HTML]{32CB00}51.82 & \color[HTML]{32CB00}35.09 & \color[HTML]{32CB00}46.31 & N/A \\
adder 16bit & 187.05 & 202.13 & 19.62 & 0.23 & 5.77 & N/A & \color[HTML]{32CB00}58.70 & \color[HTML]{32CB00}59.11 & \color[HTML]{32CB00}57.03 & \color[HTML]{32CB00}61.21 & \color[HTML]{32CB00}58.40 & N/A \\
adder 32bit & 455.43 & 513.09 & 40.54 & 0.37 & 3.37 & N/A & \color[HTML]{32CB00}21.43 & \color[HTML]{32CB00}21.43 & \color[HTML]{32CB00}22.08 & \color[HTML]{32CB00}24.32 & \color[HTML]{32CB00}41.54 & N/A \\
adder 64bit & 2464.94 & 2781.26 & 95.12 & 2.99 & 2.6 & -2.32 & \color[HTML]{32CB00}58.9 & \color[HTML]{32CB00}58.35 & \color[HTML]{32CB00}12.81 & \color[HTML]{32CB00}58.86 & \color[HTML]{32CB00}16.15 & \color[HTML]{32CB00}17.24 \\
multi 8bit & 657.47 & 757.24 & 61.36 & 0.67 & 4.84 & N/A & \color[HTML]{32CB00}1.70 & \color[HTML]{32CB00}3.96 & \color[HTML]{32CB00}28.22 & \color[HTML]{FE0000}-2.97 & \color[HTML]{32CB00}63.84 & N/A \\
multi 16bit & 2702.82 & 3209.4 & 455.11 & 2.89 & 3.51 & -3.22 & \color[HTML]{32CB00}11.85 & \color[HTML]{32CB00}12.38 & \color[HTML]{32CB00}33.08 & \color[HTML]{32CB00}10.38 & \color[HTML]{32CB00}36.18 & \color[HTML]{32CB00}39.44 \\
multi pipe 4bit & 301.41 & 341.59 & 75.75 & 0.32 & 0.48 & -0.2 & \color[HTML]{32CB00}18.8 & \color[HTML]{32CB00}22.4 & \color[HTML]{32CB00}21.83 &\color[HTML]{32CB00} 3.43 & \color[HTML]{32CB00}31.25 & \color[HTML]{32CB00}90 \\
multi pipe 8bit & 938.55 & 1078.35 & 152.57 & 0.99 & 4.26 & -3.97 & \color[HTML]{32CB00}7.311 & \color[HTML]{32CB00}8.83 & \color[HTML]{32CB00}18.94 & \color[HTML]{FE0000}-5.84 & \color[HTML]{32CB00}3.05 & \color[HTML]{32CB00}3.27 \\
div 16bit & 1554.6 & 1905.1 & 241.55 & 1.35 & 29.11 & N/A & \color[HTML]{32CB00} 37.37 & \color[HTML]{32CB00}39.32 & \color[HTML]{32CB00}3.20 & \color[HTML]{32CB00}12.59 & \color[HTML]{32CB00}73.23 & N/A \\
JC counter & 467.12 & 537.8 & 17.61 & 0.65 & 0.51 & N/A & \color[HTML]{32CB00}75.89 & \color[HTML]{32CB00}75.84 & \color[HTML]{32CB00}87.96 & \color[HTML]{32CB00}76.87 & \color[HTML]{32CB00}13.72 & N/A \\
right shifter & 61.76 & 64.61 & 32.84 & 0.101 & 0.33 & -0.02 & 0 & 0 & \color[HTML]{32CB00}11.23 & \color[HTML]{32CB00}8.91 & 0 & \color[HTML]{FE0000}-50 \\
mux & 126.06 & 133.26 & 4.03 & 0.16 & 0.46 & N/A & \color[HTML]{32CB00}25 & \color[HTML]{32CB00}25.31 & \color[HTML]{32CB00}27.29 & \color[HTML]{32CB00}20.80 & \color[HTML]{32CB00}17.39 & N/A \\
counter 12 & 54.13 & 57.48 & 18.1 & 0.07 & 0.85 & -0.59 & \color[HTML]{32CB00}10.33 & \color[HTML]{32CB00}10.39 & \color[HTML]{32CB00}32.21 & \color[HTML]{32CB00}6.94 & \color[HTML]{32CB00}7.06 & \color[HTML]{32CB00}11.86 \\
freq div & 165.19 & 175.03 & 3.77 & 0.23 & 0.39 & N/A & \color[HTML]{32CB00}22.61 & \color[HTML]{32CB00}22.63 & \color[HTML]{32CB00}27.85 & \color[HTML]{32CB00}22.02 & \color[HTML]{32CB00}2.56 & N/A \\
signal generator & 128.6 & 140.52 & 2.19 & 0.12 & 0.53 & N/A & \color[HTML]{32CB00}1.38 & \color[HTML]{32CB00}3.18 & \color[HTML]{32CB00}12.78 & \color[HTML]{FE0000}-3.36 & \color[HTML]{32CB00}13.20 & N/A \\
serial2parallel & 210.43 & 245.68 & 72.2 & 0.27 & 0.91 & -0.64 & \color[HTML]{32CB00}33.57 & \color[HTML]{32CB00}38.88 & \color[HTML]{32CB00}33.88 & \color[HTML]{32CB00}37.04 & \color[HTML]{32CB00}17.58 & \color[HTML]{32CB00}23.44 \\
edge detect & 28.72 & 30.28 & 11.01 & 0.04 & 0.48 & -0.18 & \color[HTML]{32CB00}54.87 & \color[HTML]{32CB00}54.26 & \color[HTML]{32CB00}64.12 & \color[HTML]{32CB00}60.25 & \color[HTML]{32CB00}52.08 & 0 \\
FSM & 75.73 & 80.09 & 12.91 & 0.08 & 0.86 & -0.58 & \color[HTML]{32CB00}29.52 & \color[HTML]{32CB00}27.89 & \color[HTML]{32CB00}85.51 & \color[HTML]{32CB00}25.30 & \color[HTML]{32CB00}12.79 & \color[HTML]{32CB00}22.41 \\
width 8to16 & 323.53 & 383.27 & 103.1 & 0.37 & 0.85 & -0.57 & \color[HTML]{32CB00}15.39 & \color[HTML]{32CB00}16.46 & \color[HTML]{32CB00}5.04 & \color[HTML]{32CB00}7.24 & \color[HTML]{32CB00}1.17 & \color[HTML]{32CB00}8.77 \\
calendar & 273.71 & 319.58 & 5.02 & 0.29 & 0.51 & N/A & \color[HTML]{32CB00}6.96 & \color[HTML]{32CB00}7.09 & \color[HTML]{32CB00}40.24 & 0 & 0 & N/A \\
RAM & 805.13 & 953.68 & 199.37 & 0.92 & 0.81 & -0.51 & \color[HTML]{32CB00}11.64 & \color[HTML]{32CB00}13.28 & \color[HTML]{32CB00}11.62 & \color[HTML]{32CB00}7.44 & 0 & 0 \\
asyn fifo & 2119.31 & 2487.06 & 39.73 & 2.58 & 1.16 & N/A & \color[HTML]{32CB00}1.29 & \color[HTML]{32CB00}1.16 & \color[HTML]{32CB00}4.12 & \color[HTML]{32CB00}1.55 & \color[HTML]{32CB00}59.48 & N/A \\
ALU & 3275.41 & 3394.05 & 235.8 & 2.75 & 9.81 & N/A & \color[HTML]{32CB00}29.73 & \color[HTML]{32CB00}31.35 & \color[HTML]{32CB00}36.70 & \color[HTML]{32CB00}34.18 & \color[HTML]{32CB00}1.94 & 0 \\
PE & 5065.34 & 5904.65 & 1.45 & 0.006 & 2.58 & -2.26 & \color[HTML]{32CB00}2.16 & \color[HTML]{32CB00}0.23 & \color[HTML]{32CB00}2.07 & \color[HTML]{FE0000}-0.16 & \color[HTML]{32CB00}17.44 & \color[HTML]{32CB00}19.47\\
\bottomrule
\end{tabular}%
}
\end{table*}

In Table~\ref{tab:ppa_comparison} we present our results towards high-quality PPA-aware code generation by \emph{VeriOpt}. \emph{VeriOpt} applies various PPA optimization techniques on the baseline functionally correct codes developed from multi-role LLMs (discussed in Sec.~\ref{subsec:multi_role_LLM} and results in Sec.~\ref{subsec:rq1_results}) and generates power, performance and area optimized designs. Table~\ref{tab:ppa_comparison} shows that our framework achieves significant improvement across all the metrics for every design from RTLLM~\cite{lu2024rtllm} benchmark. It improved up to $75.89\%$ for cell area, $75.84\%$ for design area, $87.96\%$ for dynamic power, and $76.87\%$ for leakage power in JC counter, up to $73.23\%$ for critical path length in div $16$-bit and $39.44\%$ for critical path slack in multi $16$-bit designs. In cases where the baseline design is a fully combinational circuit, critical path slack is marked as `N/A - Not Applicable’. \emph{VeriOpt} could not obtain improvement in a few cases where the design was already optimized, for example, in the right shifter`s area and the calendar's leakage power. There are a few negative values in the table due to the trade-offs in the nature of the design. For instance, while decreasing dynamic power, leakage power increased in the multi pipe $8$-bit and signal generator. A $50\%$ deterioration in the right shifter's critical path slack is due to its marginal primary slack of $-0.02$.

\begin{figure}[t]
    \centering
    \includegraphics[width=0.5\textwidth]{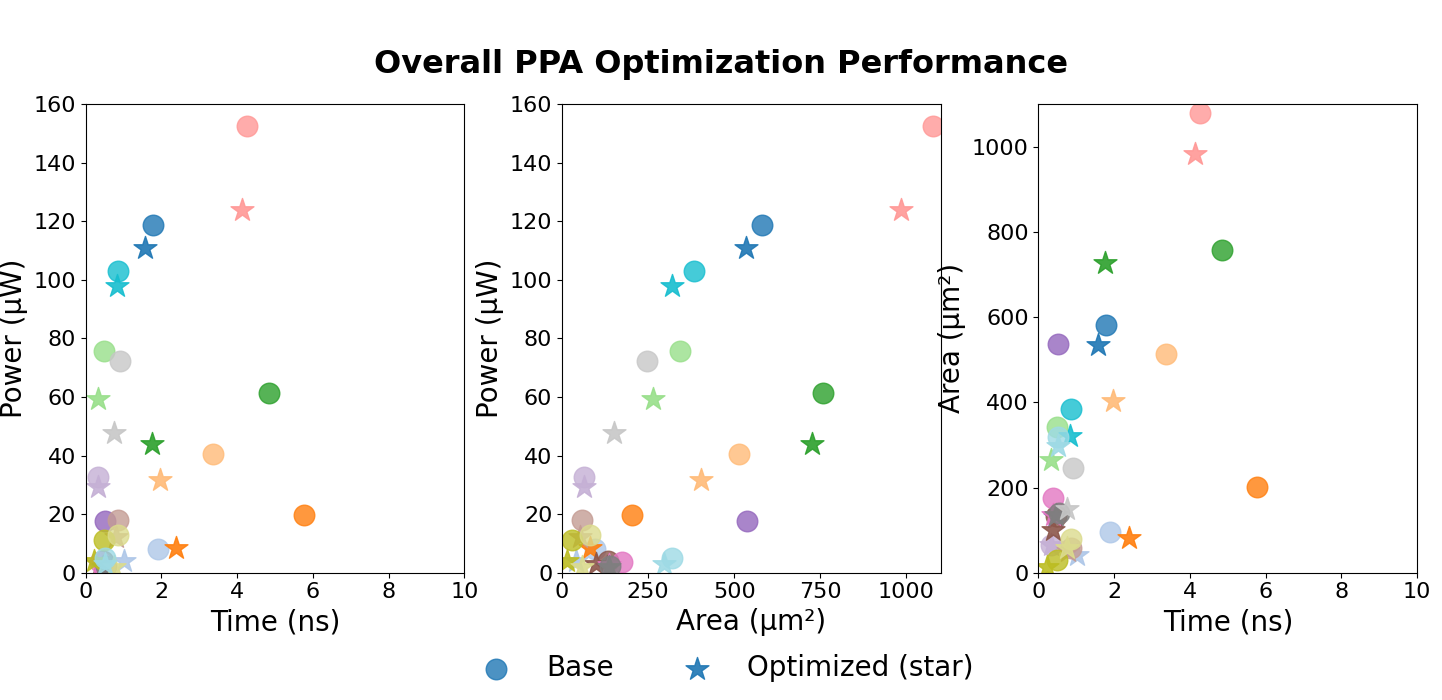}
    \caption{Overall improvement in circuit performance considering PPA trade-offs between baseline designs and \emph{VeriOpt} optimized designs}
    \label{fig:PPA plots}
\end{figure}

Fig.~\ref{fig:PPA plots} is showing the overall improvement in circuit performance considering different PPA trade-offs. \emph{VeriOpt} optimized designs achieves improvement in almost all the metrics to reduce power, area and timing which resulted in overall improvement of the design. For all the trade offs among power-vs-time, power-vs-area and area-vs-time in Fig.~\ref{fig:PPA plots}, the optimized designs achieved lower values than the baseline designs. Which indicates the designer's goal to reduce power, area and delay is achieved since the graph is showing the optimized designs are tending to zero compared to their baseline designs. These results reflect the efficiency of \emph{VeriOpt} in PPA-aware code generation. 1) It can successfully optimize the design in terms of power, performance, and area metrics by considering different trade-offs according to design constraints. And suggests the best techniques to the designer at RT-level by analyzing the design structure. 2) The in-context learning with appropriate documents in high-level natural language can make LLM significantly capable of generating high-quality RTL codes. Our results justify the exploration of \textbf{RQ2}.
\vspace{-0.1in}
\section{Discussion and Future Directions}
This section covers the effectiveness of \emph{VeriOpt}, the improvements, trade-offs, and the limitations of this work.

\vspace{-0.1in}
\subsection{Effectiveness of \emph{VeriOpt}}
Our results indicate that \emph{VeriOpt} overcomes the challenges of functionality and PPA optimization in generating Verilog RTL code by an LLM. It generates functional RTL code with high success rate demonstrating its strong potential in automated hardware generation. In terms of PPA awareness, \emph{VeriOpt} is able to generate optimized code for nearly all designs, confirming that it is not only capable of producing functionally correct RTL but also of tailoring the code for specific optimization goals in power, performance, or area, a vital metric for hardware design. This dual capability makes it a valuable tool for hardware design automation.

\subsection{Improvements and Trade Offs}
\emph{VeriOpt} is already effective in RTL code generation and PPA optimization, however, potential future improvements include fine-tuning the base LLM on domain-specific RTL datasets and implementing Retrieval-Augmented Generation (RAG). RAG would allow the system to dynamically fetch relevant design and optimization knowledge without overloading the prompt, reducing memory usage during inference time, and improving contextual consistency across iterations. Despite these improvements, trade-offs between PPA metrics are often inevitable. \emph{VeriOpt} minimizes these compromises by making intelligent optimization decisions. For example, while reducing area typically comes at the cost of performance, \emph{VeriOpt} attempts to preserve performance unless explicitly prompted by the designer.

\subsection{Limitations}
Some limitations of \emph{VeriOpt} are that models lack real-time access to synthesis tools, meaning they cannot autonomously validate their optimizations against actual timing or power reports. Instead, all improvements must be verified externally using tools like Design Compiler, introducing a manual feedback loop. Additionally, current LLMs have fixed context lengths, which makes it difficult to manage larger designs or track iterative changes effectively. So, fine-tuning and RAG could help with this limitation, so the LLM would not have to retain all this information. However, both fine-tuning and RAG require high-quality RTL code datasets with annotated PPA reports.

\section{Conclusion}
This work presented \emph{VeriOpt}, a novel framework that bridges the critical gap between functional correctness and PPA optimization in LLM-generation Verilog code. By leveraging a multi-role prompting strategy where LLM act as planner, programmer, reviewer, and evaluator, \emph{VeriOpt} ensures transparent, debug-friendly, and functionally correct RTL code generation. Unlike existing approaches that focus solely on functional correctness, \emph{VeriOpt} integrates domain-specific optimization techniques into the LLM workflow, enabling code generation that meets industrial standards for power, performance, and area (PPA). Our experiments across RTLLM benchmarks demonstrate that VeriOpt outperforms state-of-the-art baselines in both functional success and PPA efficiency, with up to $88\%$ power reduction, $76\%$ area saving, and $73\%$ timing improvement. These results underscore the potential of structured LLM workflows in automating and optimizing hardware design. Future work will explore integrating RAG and fine-tuning to further enhance \emph{VeriOpt}'s capabilities for complex system designs. By addressing the limitations of current LLMs in functionally correct and PPA efficient code generation, \emph{VeriOpt} paves the way for reliable, AI-driven EDA tools in production environments.

\newpage
\scriptsize{
\bibliographystyle{IEEEtran}
\bibliography{refs}

\begin{thebibliography}{10}
\providecommand{\url}[1]{#1}
\csname url@samestyle\endcsname
\providecommand{\newblock}{\relax}
\providecommand{\bibinfo}[2]{#2}
\providecommand{\BIBentrySTDinterwordspacing}{\spaceskip=0pt\relax}
\providecommand{\BIBentryALTinterwordstretchfactor}{4}
\providecommand{\BIBentryALTinterwordspacing}{\spaceskip=\fontdimen2\font plus
\BIBentryALTinterwordstretchfactor\fontdimen3\font minus
  \fontdimen4\font\relax}
\providecommand{\BIBforeignlanguage}[2]{{%
\expandafter\ifx\csname l@#1\endcsname\relax
\typeout{** WARNING: IEEEtran.bst: No hyphenation pattern has been}%
\typeout{** loaded for the language `#1'. Using the pattern for}%
\typeout{** the default language instead.}%
\else
\language=\csname l@#1\endcsname
\fi
#2}}
\providecommand{\BIBdecl}{\relax}
\BIBdecl

\bibitem{thakur2024verigen}
S.~Thakur, B.~Ahmad, H.~Pearce, B.~Tan, B.~Dolan-Gavitt, R.~Karri, and S.~Garg,
  ``Verigen: A large language model for verilog code generation,'' \emph{ACM
  Transactions on Design Automation of Electronic Systems}, vol.~29, no.~3, pp.
  1--31, 2024.

\bibitem{thakur2023benchmarking}
S.~Thakur, B.~Ahmad, Z.~Fan, H.~Pearce, B.~Tan, R.~Karri, B.~Dolan-Gavitt, and
  S.~Garg, ``Benchmarking large language models for automated verilog rtl code
  generation,'' in \emph{2023 Design, Automation \& Test in Europe Conference
  \& Exhibition (DATE)}.\hskip 1em plus 0.5em minus 0.4em\relax IEEE, 2023, pp.
  1--6.

\bibitem{lu2024rtllm}
Y.~Lu, S.~Liu, Q.~Zhang, and Z.~Xie, ``Rtllm: An open-source benchmark for
  design rtl generation with large language model,'' in \emph{2024 29th Asia
  and South Pacific Design Automation Conference (ASP-DAC)}.\hskip 1em plus
  0.5em minus 0.4em\relax IEEE, 2024, pp. 722--727.

\bibitem{chang2023chipgpt}
K.~Chang, Y.~Wang, H.~Ren, M.~Wang, S.~Liang, Y.~Han, H.~Li, and X.~Li,
  ``Chipgpt: How far are we from natural language hardware design,''
  \emph{arXiv preprint arXiv:2305.14019}, 2023.

\bibitem{liu2023rtlcoder}
S.~Liu, W.~Fang, Y.~Lu, Q.~Zhang, H.~Zhang, and Z.~Xie, ``Rtlcoder:
  Outperforming gpt-3.5 in design rtl generation with our open-source dataset
  and lightweight solution,'' \emph{arXiv preprint arXiv:2312.08617}, 2023.

\bibitem{tasnia2025opl4gpt}
K.~Tasnia and S.~Rahman, ``Opl4gpt: An application space exploration of optimal
  programming language for hardware design by llm,'' in \emph{Proceedings of
  the 30th Asia and South Pacific Design Automation Conference}, 2025, pp.
  981--987.

\bibitem{xu2024meic}
K.~Xu, J.~Sun, Y.~Hu, X.~Fang, W.~Shan, X.~Wang, and Z.~Jiang, ``Meic:
  Re-thinking rtl debug automation using llms,'' \emph{arXiv preprint
  arXiv:2405.06840}, 2024.

\bibitem{pei2024betterv}
Z.~Pei, H.-L. Zhen, M.~Yuan, Y.~Huang, and B.~Yu, ``Betterv: Controlled verilog
  generation with discriminative guidance,'' \emph{arXiv preprint
  arXiv:2402.03375}, 2024.

\bibitem{liu2023chipnemo}
M.~Liu, et~al., ``Chipnemo: Domain-adapted llms for chip design,'' \emph{arXiv
  preprint arXiv:2311.00176}, 2023.

\bibitem{tsai2023rtlfixer}
Y.~Tsai, M.~Liu, and H.~Ren, ``Rtlfixer: Automatically fixing rtl syntax errors
  with large language models,'' \emph{arXiv preprint arXiv:2311.16543}, 2023.

\bibitem{qiu2024explaining}
S.~Qiu, B.~Tan, and H.~Pearce, ``Explaining eda synthesis errors with llms,''
  \emph{arXiv preprint arXiv:2404.07235}, 2024.

\bibitem{zhong2023llm4eda}
R.~Zhong, X.~Du, S.~Kai, Z.~Tang, S.~Xu, H.-L. Zhen, J.~Hao, Q.~Xu, M.~Yuan,
  and J.~Yan, ``Llm4eda: Emerging progress in large language models for
  electronic design automation,'' \emph{arXiv preprint arXiv:2401.12224}, 2023.

\bibitem{yao2024rtlrewriter}
X.~Yao, Y.~Wang, X.~Li, Y.~Lian, R.~Chen, L.~Chen, M.~Yuan, H.~Xu, and B.~Yu,
  ``Rtlrewriter: Methodologies for large models aided rtl code optimization,''
  in \emph{Proceedings of the 43rd IEEE/ACM International Conference on
  Computer-Aided Design}, 2024, pp. 1--7.

\bibitem{wei2025vflow}
Y.~Wei, Z.~Huang, H.~Li, W.~W. Xing, T.-J. Lin, and L.~He, ``Vflow: Discovering
  optimal agentic workflows for verilog generation,'' \emph{arXiv preprint
  arXiv:2504.03723}, 2025.

\bibitem{zhang2024mg}
Y.~Zhang, Z.~Yu, Y.~Fu, C.~Wan, and Y.~C. Lin, ``Mg-verilog: Multi-grained
  dataset towards enhanced llm-assisted verilog generation,'' in \emph{2024
  IEEE LLM Aided Design Workshop (LAD)}.\hskip 1em plus 0.5em minus 0.4em\relax
  IEEE, 2024, pp. 1--5.

\bibitem{nakkab2024rome}
A.~Nakkab, S.~Q. Zhang, R.~Karri, and S.~Garg, ``Rome was not built in a single
  step: Hierarchical prompting for llm-based chip design,'' in
  \emph{Proceedings of the 2024 ACM/IEEE International Symposium on Machine
  Learning for CAD}, 2024, pp. 1--11.

\bibitem{sun2025paradigm}
W.~Sun, B.~Li, G.~L. Zhang, X.~Yin, C.~Zhuo, and U.~Schlichtmann,
  ``Paradigm-based automatic hdl code generation using llms,'' \emph{arXiv
  preprint arXiv:2501.12702}, 2025.

\bibitem{wang2025rtlsquad}
B.~Wang, Q.~Xiong, Z.~Xiang, L.~Wang, and R.~Chen, ``Rtlsquad: Multi-agent
  based interpretable rtl design,'' \emph{arXiv preprint arXiv:2501.05470},
  2025.

\bibitem{ho2024verilogcoder}
C.-T. Ho, H.~Ren, and B.~Khailany, ``Verilogcoder: Autonomous verilog coding
  agents with graph-based planning and abstract syntax tree (ast)-based
  waveform tracing tool,'' \emph{arXiv preprint arXiv:2408.08927}, 2024.

\bibitem{chang2024data}
K.~Chang, K.~Wang, N.~Yang, Y.~Wang, D.~Jin, W.~Zhu, Z.~Chen, C.~Li, H.~Yan,
  Y.~Zhou \emph{et~al.}, ``Data is all you need: Finetuning llms for chip
  design via an automated design-data augmentation framework,'' in
  \emph{Proceedings of the 61st ACM/IEEE Design Automation Conference}, 2024,
  pp. 1--6.

\bibitem{yang2025haven}
Y.~Yang, F.~Teng, P.~Liu, M.~Qi, C.~Lv, J.~Li, X.~Zhang, and Z.~He, ``Haven:
  Hallucination-mitigated llm for verilog code generation aligned with hdl
  engineers,'' \emph{arXiv preprint arXiv:2501.04908}, 2025.

\bibitem{guo2025survey}
C.~Guo, F.~Cheng, Z.~Du, J.~Kiessling, J.~Ku, S.~Li, Z.~Li, M.~Ma,
  T.~Molom-Ochir, B.~Morris \emph{et~al.}, ``A survey: Collaborative hardware
  and software design in the era of large language models,'' \emph{IEEE
  Circuits and Systems Magazine}, vol.~25, no.~1, pp. 35--57, 2025.

\bibitem{fang2025survey}
W.~Fang, J.~Wang, Y.~Lu, S.~Liu, Y.~Wu, Y.~Ma, and Z.~Xie, ``A survey of
  circuit foundation model: Foundation ai models for vlsi circuit design and
  eda,'' \emph{arXiv preprint arXiv:2504.03711}, 2025.

\bibitem{chen2024large}
L.~Chen, Y.~Chen, Z.~Chu, W.~Fang, T.-Y. Ho, R.~Huang, Y.~Huang, S.~Khan,
  M.~Li, X.~Li \emph{et~al.}, ``Large circuit models: opportunities and
  challenges,'' \emph{Science China Information Sciences}, vol.~67, no.~10, p.
  200402, 2024.

\bibitem{gandham2024circuitseer}
S.~Gandham, J.~Walston, S.~Samanta, L.~Yin, H.~Zheng, M.~Lin, and
  S.~Diamantidis, ``Circuitseer: Rtl post-pnr delay prediction via coupling
  functional and structural representation,'' in \emph{Proceedings of the 43rd
  IEEE/ACM International Conference on Computer-Aided Design}, 2024, pp. 1--9.

\bibitem{akyash2025simeval}
M.~Akyash and H.~Mardani~Kamali, ``Simeval: Investigating the similarity
  obstacle in llm-based hardware code generation,'' in \emph{Proceedings of the
  30th Asia and South Pacific Design Automation Conference}, 2025, pp.
  1002--1007.

\bibitem{delorenzo2024make}
M.~DeLorenzo, A.~B. Chowdhury, V.~Gohil, S.~Thakur, R.~Karri, S.~Garg, and
  J.~Rajendran, ``Make every move count: Llm-based high-quality rtl code
  generation using mcts,'' \emph{arXiv preprint arXiv:2402.03289}, 2024.

\bibitem{tang2024hivegen}
J.~Tang, J.~Qin, K.~Thorat, C.~Zhu-Tian, Y.~Cao, C.~Ding \emph{et~al.},
  ``Hivegen--hierarchical llm-based verilog generation for scalable chip
  design,'' \emph{arXiv preprint arXiv:2412.05393}, 2024.

\bibitem{weste2015cmos}
N.~H. Weste and D.~Harris, \emph{CMOS VLSI design: a circuits and systems
  perspective}.\hskip 1em plus 0.5em minus 0.4em\relax Pearson Education India,
  2015.

\bibitem{hennessy2011computer}
J.~L. Hennessy and D.~A. Patterson, \emph{Computer architecture: a quantitative
  approach}.\hskip 1em plus 0.5em minus 0.4em\relax Elsevier, 2011.

\bibitem{leiserson1991retiming}
C.~E. Leiserson and J.~B. Saxe, ``Retiming synchronous circuitry,''
  \emph{Algorithmica}, vol.~6, no.~1, pp. 5--35, 1991.

\bibitem{fu2024generalize}
W.~Fu, S.~Li, Y.~Zhao, K.~Yang, X.~Zhang, Y.~Jin, and X.~Guo, ``A generalize
  hardware debugging approach for large language models semi-syntectic
  datasets,'' \emph{Authorea Preprints}, 2024.

\bibitem{wei2022chain}
J.~Wei, X.~Wang, D.~Schuurmans, M.~Bosma, F.~Xia, E.~Chi, Q.~V. Le, D.~Zhou
  \emph{et~al.}, ``Chain-of-thought prompting elicits reasoning in large
  language models,'' \emph{Advances in neural information processing systems},
  vol.~35, pp. 24\,824--24\,837, 2022.

\bibitem{openmsp}
{A synthesizable 16bit microcontroller core},
  ``https://github.com/olgirard/openmsp430/\\tree/master/core/rtl/verilog,''
  openmsp430, 2018.

\bibitem{li2024s}
B.~Li, Z.~Di, Y.~Yang, H.~Qian, P.~Yang, H.~Hao, K.~Tang, and A.~Zhou, ``It's
  morphing time: Unleashing the potential of multiple llms via multi-objective
  optimization,'' \emph{arXiv preprint arXiv:2407.00487}, 2024.

\bibitem{li2023large}
J.~Li, C.~Tao, J.~Li, G.~Li, Z.~Jin, H.~Zhang, Z.~Fang, and F.~Liu, ``Large
  language model-aware in-context learning for code generation,'' \emph{ACM
  Transactions on Software Engineering and Methodology}, 2023.

\bibitem{loopunrolling}
{Intel®High Level Synthesis Compiler Pro Edition: Best Practices Guide},
  ``{https://www.intel.com/content/www/us/en/docs/programmable/683152/21-3/example-loop-pipelining-and-unrolling.html},''
  Intel, 2021.

\bibitem{vivadosynth}
``{https://docs.amd.com/r/2024.1-English/ug901-vivado-synthesis},
  author={{Vivado Design Suite User Guide: Synthesis (UG901)}},
  organization={AMD}, year={2024}.''

\bibitem{powergating}
{The Ultimate Guide to Power Gating},
  ``{https://anysilicon.com/power-gating/},'' anysilicon, 2025.

\bibitem{fsmlowpower}
L.~Yuan and G.~Qu, ``Fsm re-engineering for low power state encoding,''
  \emph{Proceedings International Workshop on Logic Synthesis}, 2004.

\bibitem{clockgating}
N.~Srinivasana, N.~Prakasha, S.~S. Lakshmi.Ga, and B.~T. Sundari, ``Power
  reduction by clock gating technique,'' \emph{Elsevier}, 2015.

\bibitem{loopoptimize}
L.~Song, K.~Kavi, and R.~Cytron, ``An unfolding-based loop optimization
  technique,'' \emph{International Workshop on Software and Compilers for
  Embedded Systems}, 2003.

\bibitem{stateencoding}
V.~Salauyou and W.~Bułatow, ``Optimized sequential state encoding methods for
  finite-state machines in field-programmable gate array implementations,''
  \emph{applied sciences}, 2024.

\end{thebibliography}
}
\end{document}